\documentstyle[psfig,aps,pra,multicol]{revtex}

\begin{document}

\title{Fluorescence control through multiple interference mechanisms}

\author{E.\ Paspalakis$^{1}$, C.H.\ Keitel$^{2,3}$ and  P.L.\ Knight$^{1}$}

\address{$^{1}$ Optics Section, Blackett Laboratory, Imperial College, 
London SW7 2BZ, U.K.}

\address{$^{2}$ Theoretische Quantendynamik, Fakult\"at f\"ur Physik, 
Universit\"at Freiburg,\\ Hermann-Herder-Str. 3, D-79104 Freiburg, Germany}

\address{$^{3}$ Institut f\"ur Theoretische Physik, Universit\"at
Innsbruck, Technikerstr. 25/2,\\ A-6020 Innsbruck, Austria} 

\date{\today}

\maketitle


\begin{abstract}

We discuss the spontaneous emission from a coherently prepared and microwave 
driven doublet of potentially closely spaced excited states to a common ground level. Multiple interference mechanisms are identified which may lead to fluorescence inhibition in well-separated regions of the spectrum or act jointly in cancelling the spontaneous emission. In addition to phase independent quantum interferences due to combined 
absorptions and emissions of driving field photons, we distinguish two competing phase dependent interference mechanisms as means of controlling the
fluorescence. The indistinguishable quantum paths may involve the
spontaneous emission from the same state of the doublet, originating from the
two different  components of the initial coherent superposition. Alternatively
the paths involve a different spontaneous photon from each of two decaying
states, necessarily with the same polarization. This makes these photons  indistinguishable in
principle within the uncertainty of the two decay rates. The
phase dependence arises for  both mechanisms because the interfering paths
differ by an unequal number of  stimulated absorptions and emissions of the
microwave field photons.  
\end{abstract}


\section{Introduction}
The study of spontaneous emission and the various means by which it may be modified 
and controlled has been an active area of quantum optics for many years. 
As fluorescence arises from the interaction of
the atomic system with the environmental modes, the most obvious mechanism for
control is to place the atoms in `coloured' (frequency-dependent)  reservoirs
\cite{coloured}. This can be achieved  by employing, for example, atoms in
micro-cavities  \cite{cqed}, or by placing them near the edge of photonic band
gaps \cite{bandgaps}. For atomic media in free space, quantum interferences
have become the most significant mechanism for modifying spontaneous emission.
This was first suggested in the early seventies when  Agarwal
\cite{agarwalbook} showed, for an initially prepared  degenerate {\bf V}-type
three-level  atom in free space, that the fluorescence spectrum is modified due
to interference and that population trapping occurs \cite{trapping}. This was
followed by a number of papers in the late 1970's and early 1980's on such
interferences \cite{others}.  Recently, with the need for efficient
fluorescence control  to enable such novel effects as lasing without inversion
\cite{lwi} and quantum information processing \cite{qc} to be realized, much
attention has  once more been focused on spontaneous emission dynamics from
multi-level atoms 
\cite{narducci,javanainen,hegerfeldt,zhu1,zhou,zhu2,agarwal}.  In a recent
article \cite{paspalakis}, two of us have proposed a two-colour coherent
`phase' control scheme  \cite{review} for controlling spontaneous emission in a
four-level atom. Phenomena such as extreme spectral narrowing, partial and
total  cancellation of fluorescence decay were predicted for specific values of
the atomic parameters and the  lasers phase difference. Moreover, the spectrum
was shown to be controlled very effectively and easily by changing the phase 
difference of the two lasers used for the excitation.

In this article,  using a three-level {\bf V}-type atom we discuss a further scheme 
which offers much promise for controlling spontaneous emission spectra using
the phase difference, this time, of two successively applied coherent fields
between the decaying excited doublet states.  The proposed scheme is related to
other schemes of coherent control, in particular the well-known `pump-dump'
scheme of atomic (molecular)  ionization (dissociation)
\cite{review}, the coherently driven three-level ${\bf 
\Lambda}$-type atom  of Martinez et al.\ \cite{martinez} and (not with standing
the involvement of a photonic band gap material) to the scheme  of Quang et al.
\cite{quang}. The main finding in this article is the identification of various
competing interference mechanisms leading to  phase sensitive means for
controlling the spontaneous emission spectra. If we suppose that only  one of
the two excited states decays and that the atom is initially prepared in that
decaying state, then phase insensitive interfering paths  arise which lead to
the cancellation of a specific fluorescence mode, as explained  by Knight
\cite{others} and Zhu et al.\cite{zhu1}. If instead both states are initially prepared
in a  coherent superposition and the driving field is weak then the
spectrum may acquire a Fano-type profile.  Furthermore, {\it phase sensitive}
interference
occurs and the spectrum can be controlled via this phase difference.  In
addition, if both excited states may decay, such that the emitted spontaneous
photons may not be distinguishable, even in  principle, then further phase
sensitive interfering paths via different spontaneous channels arise, similar
to those  pointed out within a driven ${\bf  \Lambda}$-type atom by Martinez et
al.\ \cite{martinez}. We discuss each  kind of interference separately by
identifying the appropriate paths and discuss the role of their co-existence
within a single scheme. In particular we find that both phase-dependent interference mechanisms may be destructive simultaneously and thus lead to spontaneous emission cancellation in different parts of the spectrum. In addition, fluorescence inhibition by one mechanism may also be enhanced due to the presence of the other. An alternative explanation of the phase dependent
interferences is also  given in the dressed states picture. In particular we
find, for an appropriate phase difference, that  the atom can be prepared in a
single decoupled dressed state in which total cancellation of the fluorescence
associated with the other (unpopulated) dressed state occurs.

This paper is organized as follows: in section II we present the atomic model 
and the basic dynamical equations in the bare states basis. We solve these
equations for the various situations of interest and present, in certain cases,  analytical
phase-dependent formulae for the time evolution of the population  and the 
spontaneous emission spectrum of this atom. In section III we present and 
analyze our basic results, and in particular discuss the situations where each
of the various interference mechanisms become significant and dominate. An
alternative analysis using the dressed-state basis is given in section IV.
Finally, we conclude in section V.

\section{The atomic model in bare states}
The atomic model considered here is shown in figure 1. A three level {\bf V}-type atom 
is initially prepared in a superposition of the two upper levels by a pump field such that  
\begin{equation}
|\psi (t = 0) \rangle = \sin{\theta} \mbox{\rm e}^{i \phi_{p}} |1, \, \{0\}\rangle + \cos{\theta}  |2, \, \{0\}\rangle \, , \label{super}
\end{equation}
where $\phi_{p}$ is the phase of the pump field. At time $t = 0$ this atom starts to interact with
a  microwave field of frequency $\omega_{c}$ and phase $\phi_{c}$ which couples the two upper levels. 
We allow both upper states to decay spontaneously to the lower state. Here, we only consider the case of 
a square pulse of the microwave field. This atomic system has been used as a model system for the 
proposal of the quantum-beat laser \cite{scully}. The dynamics of the system can be described
using the Schr\"{o}dinger equation.   Then, the wavefunction of the system at
time $t$ can be expressed in terms of the state vectors as  
\begin{equation}
|\psi(t)\rangle= a_{1}(t)|1, \, \{0\}\rangle + a_{2}(t)|2, \, \{0\}\rangle + \sum_{\bf k} a_{\bf k} (t) |0, \, \{{\bf k}\} \rangle,
\end{equation}
where ${\bf k}$ denotes both the momentum vector and the polarization of the 
emitted photon.
The Hamiltonian of the system in the interaction representation is given by
\begin{equation}
H_{int} = H_{field} + H_{vacuum} \, ,
\end{equation}
where 
\begin{eqnarray}
H_{field} &=& \Omega \mbox{\rm e}^{i\Delta t + i \phi_{c}}|1 \rangle \langle 2| + h.c. \, , \\
H_{vacuum} &=& \sum_{\bf k} g_{1{\bf k}} \mbox{\rm e}^{-i(\omega_{{\bf k}} - \omega_{10}) t} |1 \rangle \langle {\bf k}| + \sum_{\bf k} g_{2{\bf k}} \mbox{\rm e}^{-i(\omega_{{\bf k}} - \omega_{20}) t} |2 \rangle \langle {\bf k}|
+ h.c. \, ,
\end{eqnarray}
We substitute this Hamiltonian into the Schr\"{o}dinger equation and obtain the following set of equations after 
the rotating wave approximation (RWA) is carried out
\begin{eqnarray}
i\dot{a}_{1}(t) &=& \Omega \mbox{\rm e}^{i\Delta t + i \phi_{c}} a_{2}(t) +
\sum_{\bf k} g_{1{\bf k}} a_{\bf k} (t) \mbox{\rm e}^{-i(\omega_{{\bf k}} - \omega_{10}) t} \label{1} \, , \\
i\dot{a}_{2}(t) &=& \Omega \mbox{\rm e}^{-i\Delta t - i \phi_{c}} a_{1}(t) +
\sum_{\bf k} g_{2{\bf k}} a_{\bf k} (t) \mbox{\rm e}^{-i(\omega_{{\bf k}} - \omega_{20}) t} \label{2} \, , \\
i\dot{a}_{\bf k}(t) &=& g_{{\bf k}1} a_{1}(t) \mbox{\rm e}^{i(\omega_{{\bf k}} - \omega_{10}) t} + g_{{\bf k}2} a_{2}(t) \mbox{\rm e}^{i(\omega_{{\bf k}} - \omega_{20}) t} \label{k} \, .
\end{eqnarray}
Here, $\Omega$ is the Rabi frequency, which is considered real for convenience in our problem
and $\Delta \equiv \omega_{c} - \omega_{21}$ represents the microwave field detuning.
The notation $\omega_{ab} = \omega_{a} - \omega_{b}$ is used throughout 
this work. We proceed by performing a formal time integration of equation (\ref{k}) and substitute the result 
into equations (\ref{1}), (\ref{2}) to obtain 
\begin{eqnarray}
i\dot{a}_{1}(t) &=& \Omega \mbox{\rm e}^{i\Delta t + i \phi_{c}} a_{2}(t) -i\int^{t}_{0}dt^{\prime}a_{1} (t^{\prime})\sum_{\bf k} |g_{{\bf k}1}|^{2}  \mbox{\rm e}^{-i(\omega_{{\bf k}} - \omega_{10}) (t-t^{\prime})} \nonumber \\ &-& i\int^{t}_{0}dt^{\prime}a_{2} (t^{\prime})\sum_{\bf k} g_{1{\bf k}}g_{{\bf k}2} \mbox{\rm e}^{-i \omega_{{\bf k}} (t-t^{\prime}) + i\omega_{10}t - i\omega_{20}t^{\prime} } \label{1a} \, , \\
i\dot{a}_{2}(t) &=& \Omega \mbox{\rm e}^{-i\Delta t - i \phi_{c}} a_{1}(t) -i\int^{t}_{0}dt^{\prime}a_{2} (t^{\prime})\sum_{\bf k} |g_{{\bf k}2}|^{2}  \mbox{\rm e}^{-i(\omega_{{\bf k}} - \omega_{20}) (t-t^{\prime})} \nonumber \\ &-& i\int^{t}_{0}dt^{\prime}a_{1} (t^{\prime})\sum_{\bf k} g_{{\bf k}1} g_{2{\bf k}} \mbox{\rm e}^{-i \omega_{{\bf k}} (t-t^{\prime}) + i\omega_{20}t - i\omega_{10}t^{\prime} } \label{2a} \, . 
\end{eqnarray}
Once the Markov approximation is carried out within the Weisskopf-Wigner theory \cite{agarwalbook}, 
the above equations (\ref{1a}), (\ref{2a}) reduce to the following form
(after the assumption $\omega_{21} \ll \omega_{10}, \omega_{20}$ is taken into account)
\begin{eqnarray}
\dot{a}_{1}(t) &=& - \frac{\gamma_{1}}{2} a_{1} (t) - ( i \Omega \mbox{\rm e}^{i\Delta t + i \phi_{c}} + p \frac{\sqrt{\gamma_{1}\gamma_{2}}}{2}
\mbox{\rm e}^{- i\omega_{21}t}) a_{2} (t)   \label{1b} \, , \\
\dot{a}_{2}(t) &=& -(i \Omega \mbox{\rm e}^{-i\Delta t - i \phi_{c}} + p \frac{\sqrt{\gamma_{1}\gamma_{2}}}{2}\mbox{\rm e}^{i \omega_{21}t})  a_{1}(t)  - \frac{\gamma_{2}}{2} a_{2} (t) \label{2b} \, .
\end{eqnarray}
Here, $\gamma_{m} \equiv 2\pi |g_{{\bf k}m}|^{2}D(\omega_{m0})$ $(m = 1,2)$ and the radiative shifts, which are 
related to the Lamb shift, will be omitted in this approach. 
The term $(p\frac{\sqrt{\gamma_{1}\gamma_{2}}}{2}\mbox{\rm e}^{\pm i \omega_{21}t})$ is a common term which arises if 
quantum interference from both spontaneous emission channels from the two
closely spaced upper levels is involved \cite{agarwalbook}.  The parameter $p$
denotes the alignment of the two matrix elements and is defined as  $p \equiv
\vec{\mu}_{20} \cdot \vec{\mu}_{01} \big{/}  |\vec{\mu}_{20}||
\vec{\mu}_{01}|$.  For orthogonal matrix elements this yields $p = 0$ (no
interference) and for parallel matrix elements we obtain $p = 1$ (maximum
interference). In the language of quantum pathway interference, different
pathways involving different spontaneously emitted photons may only be
indistinguishable, even in principle, if the corresponding transitions give
rise to photons of identical polarization. This means that the associated
dipoles have to be parallel. In the language of off-diagonal couplings, those
extra terms in $p$ may only arise if both  dipoles involved interact with the
same modes of the vacuum. Since the time dependence of this term is $\mbox{\rm e}^{\pm i \omega_{21}t}$ this term can be omitted from these equations only in the case that 
the energy difference of the two upper levels $\omega_{21}$ is larger than 
the decay rates $\gamma_{1}, \gamma_{2}$ \cite{zhu2}.  However, for a moment, let us omit these
interference terms by setting $p = 0$ in order to obtain analytical solutions of
these equations. We will return to the importance of such
an interference in the calculation of the spectrum.  Thus, without the
contributions from the $p$ terms in  equations (\ref{1b}), (\ref{2b})  we
obtain \begin{eqnarray} \dot{b}_{1}(t) &=& - \frac{\gamma_{1}}{2} b_{1} (t) - i
\Omega \mbox{\rm e}^{i \phi_{c}} b_{2} (t)   \label{1c} \, , \\ \dot{b}_{2}(t)
&=& - i \Omega \mbox{\rm e}^{- i \phi_{c}} b_{1}(t)  + (i\Delta -
\frac{\gamma_{2}}{2}) b_{2} (t) \label{2c} \, , \\ 
\dot{b}_{\bf k}(t) &=& -i
g_{{\bf k}1} b_{1}(t) \mbox{\rm e}^{i(\delta_{\bf k} + \frac{1}{2}\omega_{21})t} -i
g_{{\bf k}2} b_{2}(t)  \mbox{\rm e}^{i(\delta_{\bf k} - \frac{1}{2}\omega_{21} -
\Delta)t} \label{ka} \, . 
\end{eqnarray} 
We here define $b_{1} (t) \equiv a_{1}
(t)$, $b_{2} (t) \equiv a_{2} (t)  \mbox{\rm e}^{i\Delta t}$, $b_{{\bf k}} (t)
\equiv a_{{\bf k}} (t)$ and  $\delta_{{\bf k}} \equiv \omega_{{\bf k}} -
(\omega_{10} + \omega_{20})/{2}$.

Equations (\ref{1c}), (\ref{2c}) have the following solutions 
if $\lambda_{1} \neq \lambda_{2}$,
\begin{eqnarray}
b_{1} (t) = C_{1} \mbox{\rm e}^{\lambda_{1} t} + 
C_{2} \mbox{\rm e}^{\lambda_{2} t} \quad , \quad 
b_{2} (t) = C^{\prime}_{1} \mbox{\rm e}^{\lambda_{1} t} + 
C^{\prime}_{2} \mbox{\rm e}^{\lambda_{2} t} \, , \label{sol}
\end{eqnarray}
where
\begin{eqnarray}
\lambda_{1,2} &=& \frac{i \Delta}{2} - \frac{\gamma_{1} + \gamma_{2}}{4} 
\pm \frac{i}{2} \sqrt{ 4 \Big{[} \Omega^{2} - (i \Delta - \frac{\gamma_{2}}{2})
\frac{\gamma_{1}}{2} \Big{]} - (i\Delta - \frac{\gamma_{1} + \gamma_{2}}{2})^{2} } \, , \\
C_{1} &=& \frac{1}{\lambda_{2} - \lambda_{1}} \Big{[} (\lambda_{2} + \frac{\gamma_{1}}{2}) \sin{\theta} + i \Omega 
\mbox{\rm e}^{i \delta \phi} \cos{\theta}  \Big{]} \, , \\
C_{2} &=& \frac{1}{\lambda_{1} - \lambda_{2}} \Big{[} (\lambda_{1} + \frac{\gamma_{1}}{2}) \sin{\theta} + i \Omega 
\mbox{\rm e}^{i \delta \phi} \cos{\theta}  \Big{]} \, , \\
C^{\prime}_{1} &=& \frac{1}{\lambda_{2} - \lambda_{1}} \Big{[} (\lambda_{2} + \frac{\gamma_{2}}{2} - i\Delta) \cos{\theta} + i \Omega 
\mbox{\rm e}^{-i \delta \phi} \sin{\theta}  \Big{]} \, , \\
C^{\prime}_{2} &=& \frac{1}{\lambda_{1} - \lambda_{2}} \Big{[} (\lambda_{1} + \frac{\gamma_{2}}{2} - i\Delta) \cos{\theta} + i \Omega 
\mbox{\rm e}^{-i \delta \phi} \sin{\theta}  \Big{]} \, .
\end{eqnarray}
Here we define the relative phase between the pump and coupling fields as $\delta\phi \equiv \phi_{c} - \phi_{p}$, which as we will show later has a crucial role in the behaviour of the system. Also, for convenience, we have 
set arbitrarily the pump field's phase to zero. So, the above phase
difference simply reduces to the phase of the coupling field.
Using equation (\ref{ka}) and solutions (\ref{sol}) we obtain
\begin{eqnarray}
b_{\bf k}(t) &=& -\frac{g_{{\bf k}1} C_{1}}{\delta_{{\bf k}} + \frac{\omega_{21}}{2} -
i\lambda_{1}} \Big{[} \mbox{\rm e}^{i (\delta_{{\bf k}} + \frac{\omega_{21}}{2} - i\lambda_{1})t} - 1 \Big{]} \nonumber \\
&-& \frac{g_{{\bf k}1}C_{2}}{\delta_{{\bf k}} + \frac{\omega_{21}}{2} -
i\lambda_{2}} \Big{[} \mbox{\rm e}^{i (\delta_{{\bf k}} + \frac{\omega_{21}}{2} - i\lambda_{2})t} - 1 \Big{]} \nonumber \\
&-& \frac{g_{{\bf k}2}C^{\prime}_{1}}{\delta_{{\bf k}} - \frac{\omega_{21}}{2} -
\Delta - i\lambda_{1}} \Big{[} \mbox{\rm e}^{i (\delta_{{\bf k}} - \frac{\omega_{21}}{2} - \Delta - i\lambda_{1})t} - 1 \Big{]} \nonumber \\
&-& \frac{g_{{\bf k}2}C^{\prime}_{2}}{\delta_{{\bf k}} - \frac{\omega_{21}}{2} -
\Delta - i\lambda_{2}} \Big{[} \mbox{\rm e}^{i (\delta_{{\bf k}} - \frac{\omega_{21}}{2} - \Delta - i\lambda_{2})t} - 1 \Big{]} \, ,
\end{eqnarray}
which in the limit $t \rightarrow \infty$ yields
\begin{eqnarray}
b_{\bf k}(\infty) &=& g_{{\bf k}1} \bigg{[} \frac{C_{1}}{\delta_{{\bf k}} + \frac{\omega_{21}}{2} - i\lambda_{1}}
 + \frac{C_{2}}{\delta_{{\bf k}} + \frac{\omega_{21}}{2} -
i\lambda_{2}} \bigg{]} \nonumber \\ &+& g_{{\bf k}2} \bigg{[} \frac{C^{\prime}_{1}}{\delta_{{\bf k}} - \frac{\omega_{21}}{2} - \Delta - i\lambda_{1}}
 + \frac{C^{\prime}_{2}}{\delta_{{\bf k}} - \frac{\omega_{21}}{2} - \Delta -
i\lambda_{2}} \bigg{]} \, \label{ak}. 
\end{eqnarray}
The spontaneous emission spectrum $S(\delta_{{\bf k}})$ in this model
will be proportional to $|b_{\bf k}(\infty)|^{2}$ \cite{zhu2},
\begin{eqnarray}
S(\delta_{{\bf k}}) &\sim& \gamma_{1} \bigg{|} \frac{C_{1}}{\delta_{{\bf k}} + \frac{\omega_{21}}{2} - i\lambda_{1}}
 + \frac{C_{2}}{\delta_{{\bf k}} + \frac{\omega_{21}}{2} -
i\lambda_{2}} \bigg{|}^{2} \nonumber \\ 
&+& \gamma_{2} \bigg{|} \frac{C^{\prime}_{1}}{\delta_{{\bf k}} - \frac{\omega_{21}}{2} - \Delta - i\lambda_{1}}
 + \frac{C^{\prime}_{2}}{\delta_{{\bf k}} - \frac{\omega_{21}}{2} - \Delta -
i\lambda_{2}} \bigg{|}^{2} \, \label{spec} .
\end{eqnarray}

By inspection of formula (\ref{spec}) we identify various sources of quantum interference
which we would like to associate with the additional indistinguishable paths shown in 
figure 2. We consider in figure 2(a) the direct spontaneous emission from, for example, the 
lower state $|1 \rangle$. Even with $p=\gamma_2=\cos \theta =0$, an expansion in $\Omega^2 $ of the spectrum would give rise to extra paths due to combined 
absorption and emission of stimulated driving field photons, where the leading 
term is depicted in figure  2(b1). In fact the $\Omega^2$ terms in $\lambda_{1,2}$ should 
be read as $ \Omega {\rm e}^{{\rm i} \phi_c} \Omega {\rm e}^{ - {\rm i} \phi_c}$, showing that here 
an absorption of a microwave photon is always associated with the subsequent emission of another,
cancelling the phase dependence. 
Furthermore, for $\cos \theta \ne 0$, thus initial population also in the upper 
excited state, we obtain new phase--dependent terms in $C_1$ and $C_2$ corresponding
to interfering paths as depicted in figure 2(b2). Here the path differs from that in 2(a) 
only by a single photon of the driving field, explaining the phase dependence. 

For $p \neq 0$ we should go back to the equations of motion (\ref{k}), (\ref{1b}), (\ref{2b}), and study numerically the behaviour of the system. Then, the most complex
additional phase dependent interference arise, which is associated with
paths involving indistinguishable spontaneous photons from different transitions
as depicted in figs.\ 2(a), 2(b3).  This interference becomes
maximal if both transitions to the ground state are exactly parallel and as we will 
see later, when both levels are closely spaced, i.e. photons arising from both 
transitions become essentially indistinguishable. This type of interference
has been drown much attention recently 
both for its effects on spontaneous emission dynamics of multi-level atoms \cite{hegerfeldt,zhou,zhu2,agarwal,paspalakis} but also for its effects on the absorption, dispersion and population dynamics of these atoms \cite{paspalakis2,menon,zhou2}.
Actually, it is this type of interference that leads to the phase dependence 
in the microwave driven ${\bf  \Lambda}$-type system studied by Martinez et al.\  \cite{martinez} as happens in our scheme if the system is initially in 
one of the two excited states (either $\sin{\theta}$ or $\cos{\theta} = 1$).
If our system is furthermore initially prepared  in a superposition of states $|1 \rangle$
and  $|2 \rangle$ $(\sin{\theta}, \cos{\theta} \neq 1)$ then phase dependent dynamics is observable even if $p = 0$ as pointed out earlier, due to the
processes of figs.\ 2(a) and 2(b2). 

\section{Discussion}

We begin with an analysis of the system when one of the spontaneous decay 
rates is much smaller than the other and can be omitted, say  $\gamma_{2} = 0$.
This reduces our system to that studied by
Knight \cite{others}  and by Zhu et al.\ \cite{zhu1}. 
If we choose the atom to be initially in state $|1 \rangle$ $(\sin{\theta} = 1 \, 
, \, \cos{\theta} = 0)$ then quantum interference can lead to complete or partial 
cancellation of specific emission modes and to spectral narrowing of one of the 
two peaks of the spectrum. The leading interfering process for weak driving fields is depicted in fig.\ 2(b1).
The system in this situation, however, is not phase sensitive. This occurs because stimulated 
absorption of a driving field photon is associated 
with the subsequent stimulated emission of another photon, taking it back to the decaying state 
and thus cancelling the phase dependence.  

We proceed now with the case of an initial superposition of excited states, in particular 
when $\Omega \ll \gamma_{1}$, $\theta = \pi/4$ and $\Delta = 0$. In this case we can 
derive the following formulae for our system
\begin{eqnarray}
\lambda_{1} &\approx& -\frac{\gamma_{1}}{2} + 2 \frac{\Omega^{2}}{\gamma_{1}}\, , \, \lambda_{2} \approx - 2 \frac{\Omega^{2}}{\gamma_{1}}\, ,\\
C_{1} &\approx& \frac{\sqrt{2}}{\gamma_{1}} \Big{[}
\frac{\gamma_{1}}{2} + i \Omega \mbox{\rm e}^{i \delta\phi} \Big{]} \, , \, C_{2} \approx - \frac{\sqrt{2}}{\gamma_{1}} i \Omega \mbox{\rm e}^{i \delta\phi} \, ,
\end{eqnarray}
and 
\begin{equation}
b_{\bf k}(\infty) \approx \frac{g_{{\bf k}1}(\delta_{{\bf k}} + \frac{\omega_{21}}{2} + \Omega \mbox{\rm e}^{i \delta\phi})}{\sqrt{2}(\delta_{{\bf k}} + \frac{\omega_{21}}{2} + i\frac{\gamma_{1}}{2} - 2i \frac{\Omega^{2}}{\gamma_{1}}) (\delta_{{\bf k}} + \frac{\omega_{21}}{2} + 2i \frac{\Omega^{2}}{\gamma_{1}}) } \, .
\end{equation}
An interesting feature occurs for $\delta\phi = 0$ ($\delta\phi = \pi$) since 
the spectrum becomes zero at $\delta_{{\bf k}} = - \omega_{21}/2 -\Omega$ ($\delta_{{\bf k}} =  - \omega_{21}/2 + \Omega$). 
The form of the spectrum for this case is shown in figure 3. For the particular
values of $\delta\phi = 0$ and  $\delta\phi = \pi$ the spectrum resembles a
Fano-type form \cite{fano}, as shown in figures 3(a) and 3(c).  
The zero  disappears
for other values of the phase $\delta\phi$, as for example for $\delta\phi =
\pi/2$, as shown in  figure 3(b) where the spectrum is simply a single
Lorentzian peak.    
The existence of this interference in the spectrum is associated with
additional interfering paths as  indicated in figure 2(b2). Obviously the
interferences need not be destructive for each choice of parameters, so that 
we have an exact zero in the spectrum only for particular choices in the phase
$\delta\phi$. An equally intuitive way of understanding the interferences can
be given in the dressed states picture and a  detailed explanation will be
given in the following section. A similar effect has been investigated by Coleman and Knight \cite{coleman} in the related area of resonant two-photon
ionization and by Agassi in the area of spontaneous emission of autoionizing states \cite{agassi}. We note, that the appearance of exact zeros in  the fluorescence
spectrum is very sensitive to the radiative emission of the other level $(|2
\rangle)$. Then  various interfering paths analogous to figure 2(a) and figure
2(b3) arise with decay from the upper excited state $(|2 \rangle)$, and
obviously those do not need to interference destructively for the same
parameters at the same frequency.  For example, in figure 4 we present the
results for the spontaneous emission spectrum of this atom, now assuming a very
weak emission from level $|2 \rangle$, i.e.\ $\gamma_{2} \neq 0$ but also
$\gamma_{2} \ll \gamma_{1}$.  We note that the exact zero noted earlier has
disappeared for those parameters. In this case the  spectrum will also depend
on the value of $p$. With both decay channels open and with $p \neq 0$ we
involve further interfering channels such as those presented in figure 2(b3).
The dashed curve in figure 4 shows the results with $p = 0$ using eq.\ (\ref{spec}) and the solid
curve the results with $p = 1$ from a numerical solution of eqs.\ (\ref{k}), (\ref{1b}), (\ref{2b}). For $p = 0$ we can immediately see that even for a small
$\gamma_{2}$ of $0.5 \Omega$, as shown in figure 4(a), the precise zero disappears and a minimum well above zero takes its place in the spectrum. This minimum is further lifted when the
decay rate of the upper excited state increases, as displayed in figures 4(b) and (c). In addition, a second peak appears in the spectrum 
due to the second decay rate. In the case that $p = 1$ the behaviour of the system is more complex due to the additional interfering channels originating from the mechanism displayed in figs.\ 2(a), 2(b3). Firstly, we note specifically from figs.\ 4(b) and (c) that the presence of this additional interference mechanism enhances the inhibition of spontaneous emission in the Fano minimum. Secondly, an additional zero appears in the spectrum which is solely associated with the new interference mechanism. This zero is quite stable to the increase of the decay rate $\gamma_{2}$. These results clearly show the co-existence of two distinguished mechanisms for fluorescence inhibition, which can act either independently or jointly in order to modify the spectrum.

In the following, we consider the case when $\gamma_{1} = \gamma_{2} = \gamma$ and $\Delta = 0$ and examine the phase 
dependence of the system. 
We have seen
from the previous discussion that the Fano interferences due to the decay
channels $\gamma_{1}$ and $\gamma_{2}$ do not interfere destructively at the
same frequency. Thus, for $\gamma_{1} = \gamma_{2}$, we do not expect any zeros due to Fano interference and the other phase dependent mechanism,  which is displayed in fig. 2(b3), will dominate. In this case the solutions (\ref{sol}) obtain the form
\begin{eqnarray}
b_{1} (t) & =& \frac{1}{2} \Big{[} (\sin{\theta} - \mbox{\rm e}^{i \delta\phi} \cos{\theta}) \mbox{\rm e}^{ -(\gamma/2 - i \Omega)t} + 
(\sin{\theta} + \mbox{\rm e}^{i \delta\phi} \cos{\theta}) \mbox{\rm e}^{ -(\gamma/2 + i \Omega)t} \Big{]} \, ,  \label{solb1}\\ 
b_{2} (t) & =& \frac{1}{2} \Big{[} (\cos{\theta} - \mbox{\rm e}^{-i \delta\phi} \sin{\theta}) \mbox{\rm e}^{ -(\gamma/2 - i \Omega)t} + 
(\cos{\theta} + \mbox{\rm e}^{-i \delta\phi} \sin{\theta}) \mbox{\rm e}^{ -(\gamma/2 + i \Omega)t} \Big{]} \, ,  
\label{solb2}
\end{eqnarray}
and finally equation (\ref{ak}) reads,
\begin{eqnarray}
b_{\bf k}(\infty) &=&  \frac{g_{{\bf k}1}}{2} \bigg{[} \frac{ \sin{\theta} - \mbox{\rm e}^{i \delta\phi} \cos{\theta}}{\delta_{{\bf k}} + \frac{\omega_{21}}{2} + \Omega + i \frac{\gamma}{2}}
 + \frac{\sin{\theta} + \mbox{\rm e}^{i \delta\phi} \cos{\theta}}{\delta_{{\bf k}} + \frac{\omega_{21}}{2} -
\Omega + i \frac{\gamma}{2}} \bigg{]} \nonumber \\ &+&  \frac{g_{{\bf k}2}}{2} \bigg{[} \frac{\cos{\theta} - \mbox{\rm e}^{-i \delta\phi} \sin{\theta}}{\delta_{{\bf k}} - \frac{\omega_{21}}{2} - \Delta + \Omega + i \frac{\gamma}{2}}
 + \frac{\cos{\theta} + \mbox{\rm e}^{-i \delta\phi} \sin{\theta}}{\delta_{{\bf k}} - \frac{\omega_{21}}{2} - \Delta -
\Omega + i \frac{\gamma}{2}} \bigg{]} \, \label{bk1}. 
\end{eqnarray}
The case when $\sin{\theta} = \cos{\theta} = 1/\sqrt{2}$ is once more very interesting.
When $\delta\phi = 0$ the 
above equations (\ref{solb1}), (\ref{solb2}) take the simple form $P_{m} (t) = |b_{m}(t)|^{2} = 
\frac{1}{2}\mbox{\rm e}^{-\gamma t}$, 
$m = 1,2$. This indicates a simple exponential decaying behaviour of the populations. The same result is obtained for $\delta\phi = \pi$. For $\delta\phi = 0.5\pi$ the 
populations become $P_{m} (t)= \frac{1}{2}\mbox{\rm e}^{-\gamma t} [1 \pm \sin{(2 \Omega t)}]$, $m = 1,2$. In this case populations of each state show decaying Rabi oscillations, but we should note that the total population just decays exponentially as in the case that $\delta \phi = 0, \pi$.
The spontaneous emission spectrum behaviour displays very rich features. In figure 5 we present the results
for the spontaneous emission spectrum for several values of the phase $\delta\phi$ and for both cases of
$p = 0$ (shown with dashed curves) and $p = 1$ (shown with solid curves). The spectrum is clearly double-peaked for $\delta\phi = 0$, as shown in figure 5(a), and for 
$\delta\phi = \pi$, shown in figure 5(e).
It is easy to verify from eqs. (\ref{spec}), (\ref{bk1}) that, in the case $p = 0$, the spectrum for these atomic 
parameters and phase values is given by the sum of two Lorentzian curves. This is related to the occupation 
of just a single dressed state and will be analyzed later. However,
for any other value between $0$ and $\pi$ of the relative phase $\delta\phi$ the emission spectrum is made up of four peaks (two associated with each bare state contribution for each of the two dressed states), as shown in figures 5(b)--(d).  For $p=1$ we note destructive or constructive interference between the two bare state contributions from each dressed 
state, similar to that observed in the ${\bf \Lambda}$-system \cite{martinez}. 
Total line elimination cannot arise due
to this interference mechanism. This is only possible for an appropriate initial preparation in one of the dressed
states and decoupling to the other, and naturally but trivially occurs if one of the bare channels is forbidden. 

\section{Dressed state analysis}
The dressed state approach of dynamics \cite{dressed} is a very useful tool for further understanding the behaviour
of the system as presented in the previous section. The diagonalization of the field-atom interaction Hamiltonian (in a rotating frame)
\begin{equation}
H^{\prime}_{field} = -\Delta |2 \rangle \langle 2| + \Omega_{c} \mbox{\rm e}^{i \phi_{c}}|1 \rangle \langle 2| + \Omega_{c} \mbox{\rm e}^{-i \phi_{c}}|2 \rangle \langle 1| \label{ham} \, ,
\end{equation}
lead to the dressed eigenstates
\begin{equation}
|+ \rangle = \mbox{\rm e}^{-i \phi_{c}} \cos{\Psi} |1 \rangle + \sin{\Psi} |2 \rangle \, , \,
|- \rangle = - \sin{\Psi} |1 \rangle + \mbox{\rm e}^{i \phi_{c}} \cos{\Psi} |2 \rangle \, ,
\end{equation}
where $\tan{\Psi} = {E_{-}}\big{/}{\Omega}$.

Here, $E_{\pm} \equiv (-\Delta \pm \sqrt{\Delta^{2} + 4 \Omega^{2}})/2$ are the eigenenergies of Hamiltonian
(\ref{ham}). The wavefunction of the system at time $t$ can be expressed in terms of the dressed states as
\begin{equation}
|\psi(t)\rangle= b_{+}(t)\mbox{\rm e}^{-iE_{+} t}|+, \, \{0\}\rangle + b_{-}(t)\mbox{\rm e}^{-iE_{-} t}|-, \, \{0\}\rangle + \sum_{\bf k} b_{\bf k} (t) \mbox{\rm e}^{-i\omega_{{\bf k}} t} |0, \, \{{\bf k}\} \rangle.
\end{equation}
The equations of motion of these amplitudes are given by
\begin{eqnarray}
\dot{b}_{+}(t) &=& - \frac{1}{2} \gamma_{+} b_{+}(t) - \frac{1}{2} \gamma_{\pm} b_{-}(t) \mbox{\rm e}^{i E^{\prime}_{\pm} t}  \, , \\
\dot{b}_{-}(t) &=& - \frac{1}{2} \gamma_{-} b_{-}(t) - \frac{1}{2} \gamma_{\mp} b_{+}(t) \mbox{\rm e}^{-i E^{\prime}_{\pm} t} \, . 
\end{eqnarray}
Here, $\gamma_{+}$ ($\gamma_{-}$) is the $|+ \rangle$ ($|- \rangle$) dressed state decay to the ground state and $\gamma_{\mp}$, $\gamma_{\pm}$ are relaxations among the dressed states, as it is shown in figure 1(b). Also $E^{\prime}_{\pm} = E_{+} - E_{-}$ is the energy difference of the dressed states.

The probability amplitudes of the dressed states are connected to the probability amplitudes 
of the bare states by the following relations,
\begin{equation}
b_{+}(t) = \mbox{\rm e}^{i \phi_{c}} \cos{\Psi}b_{1}(t) + \sin{\Psi}b_{2}(t) \, , \, b_{-}(t) = - \sin{\Psi}b_{1}(t) + \mbox{\rm e}^{-i \phi_{c}} \cos{\Psi}b_{2}(t)  \, . \label{dressed2}
\end{equation}
Using the initial condition, eq.\ (\ref{super}), the above equations (\ref{dressed2}) will reduce to
\begin{eqnarray}
b_{+}(t=0) &=& \mbox{\rm e}^{i \delta\phi} \cos{\Psi}\sin{\theta} + \sin{\Psi}\cos{\theta} \, , \\  
b_{-}(t=0) &=& \mbox{\rm e}^{-i \delta\phi}[- \sin{\Psi}\sin{\theta}\mbox{\rm e}^{i \delta\phi} +  \cos{\Psi}\cos{\theta}]  \, ,
\end{eqnarray}
which for the resonance case $(\Delta = 0)$  yields
\begin{equation}
b_{+}(t=0) = \frac{1}{\sqrt{2}}(\mbox{\rm e}^{i \delta\phi}\sin{\theta} - \cos{\theta}) \, , \, 
b_{-}(t=0) = \frac{\mbox{\rm e}^{-i \delta\phi}}{\sqrt{2}}[\sin{\theta}\mbox{\rm e}^{i \delta\phi} + \cos{\theta}]  \, . \label{init}
\end{equation}
So if $\theta = \pi/4$ and $\delta\phi = 0$ $(\delta\phi =\pi)$ only the $|- \rangle$ $(|+ \rangle)$ dressed 
state is populated initially. 
In the weak-field limit and on-resonance, the positions of the two dressed-states are close in energy $E^{\prime}_{\pm} \approx 0$ and the two dressed states are efficiently coupled to each other via 
the corresponding transitions to the ground state. Such a coupling is usually referred as the off-diagonal coupling of the dressed states \cite{coleman}.  
In the specific case that one of the two bare excited states decays ($\gamma_{2} = 0$), the dressed decay rates read
\begin{eqnarray}
\gamma_{+} &=& \gamma_{-} = \frac{\gamma_{1}}{2}  \, ,\\
\gamma_{\pm} &=& \gamma_{\mp} = \sqrt{\gamma_{+} \gamma_{-}} = \frac{\gamma_{1}}{2}  \, .
\end{eqnarray}
If system is initially in one of the dressed states ($|- \rangle$ 
dressed-state in the case of figure 3(a) and  $|+ \rangle$ dressed-state 
in the case of figure 3(c)) then population is efficiently transfered to the other dressed state due to this coupling via the ground state. 
This opens new channels for spontaneous emission and consequently interference
structures in the spectrum, similar to a Fano-type behaviour. For both states decaying transfer also occurs but the actual forms of the dressed decay rates is
far more complicated. 

In the strong-field limit the energies of the dressed-states differ
substantially and the off-diagonal coupling is very weak, as the oscillating terms $\mbox{\rm e}^{\pm i E^{\prime}_{\pm} t}$ become large.
In this case, the two dressed-states can be thought of decaying independently
\cite{coleman}.  Depending on the relative phase $\delta\phi$ either the
positive or the negative frequency peaks of the spectrum can completely
disappear, as can be seen in figures 5(a) and 5(e). This occurs as we can
selectively populate one of the dressed states initially by appropriately
choosing the phase difference $\delta\phi$ as explained by eq.\ (\ref{init}).
This dressed-state will subsequently decay without any transfer to the other
dressed-state and a two peak structure may be created in the spectrum, where
the two peaks are separated by the bare states doublet spacing (given both 
bares states can decay). 
Even without selective dressed state preparation there is no interference in
the strong field limit based on dressed off-diagonal couplings and extra paths
due to dressed population transfer. This would furthermore be impossible because
photons emitted from different dressed states become well distinguishable if
they are apart in frequency by more than the largest spontaneous emission rate.
In spite of a large dressed states separation however, interference may still
occur according to the mechanism described in figure 2(b3). This is because
the two bare contributions of each dressed state then represent the different
corner stones for interfering paths rather than different dressed states,
rendering only the spacing between the bare states relevant in this situation.

\section{Conclusions}

In this paper we have investigated the effects of the relative phase
between a pump and a coupling field on the spontaneous emission spectrum
from a three level {\bf V}-type system. We have shown that the modification of 
this phase difference allows us to efficiently control the shape of the 
spontaneous emission spectrum and the population dynamics. Various
interference mechanisms were identified from the underlying transition physics. 
The various phase dependent and independent interferences responsible for the
spectral structure depend sensitively on the relative energy and decay rates 
of the two excited states and on the orientation of the corresponding dipoles.
In particular, we have shown that in the case in which only one excited state
decays, the spontaneous emission spectrum can exhibit a Fano-type behaviour in
the weak-field limit and cancellation  of the emission in a specific vacuum
mode (i.e. specific radiated frequencies) occurs.  In the dressed state picture
this originates from interferences involving indistinguishable photons due to an
off-diagonal coupling between the dressed states. In this case the addition of decay to the other excited state could lead to more complicated sources of quantum interference. We have also emphasised
that the dressed states decouple from each other for strong driving fields
yielding an inhibition of this source of interference. As a consequence of this
decoupling, for specific initial preparation of solely one dressed state, we
noted the inhibition of the total fluorescence corresponding  to the other
dressed state. This is controlled using the relative phase $\delta\phi$ which
arises from the fact that the emission paths differ by an unequal number of 
stimulated absorptions and emissions of the microwave field photons. If the 
excited doublet states are closely spaced with respect to both decay rates and
the corresponding emission dipoles are parallel, a different phase dependent
source of interference arises. This mechanism also survives in the strong
driving field limit because it is not based on the coupling among the 
two dressed states as the previous mechanism but on that of the bare states contribution
for each dressed state independently. In general, all these different sources
of interference are present and compete, giving rise to a rich regime of
parameters to shape the spectrum.

An experimental realisation of the phenomena discussed in this paper is 
possible using
recent developments in quantum optics. For the first interference case,
where the dipole moments of the two spontaneous emission transitions
are orthogonal or the two upper states are well separated, the only essential
part is the preparation of the excited doublet, as the atomic configuration
can be realized in many atomic (or molecular) systems. This preparation can
be achieved, for example, by pulses of specific area or
by adiabatic transfer methods \cite{shore}. The second case is
more restrictive in finding the proper atomic (or molecular) configuration as it requires the spontaneous emission matrix elements
to be parallel, and the quantum states close in energy. However, a similar configuration has been found in sodium dimers and already evidence of a related interference
have been experimentally demonstrated \cite{xia}. Other possible experimental
proposals include the use of atomic hydrogen \cite{zhu2,zhou2}. Finally, it should be noted
that this type of interference 
has been recently observed in tunnelling transitions from semiconductor quantum wells \cite{faist,imamoglu}.

\section*{Acknowledgements}

We would like to thank M.B. Plenio for fruitful discussions. This work has been supported by the UK Engineering and Physical Sciences Research Council and the European Commission. C.H.K. acknowledges a Marie Curie Research Fellowship from the European Commission and a ``Nachwuchsgruppe'' from the German Research Council funded SFB 276.

\pagebreak

\begin{figure}
\centerline{\hbox{\psfig{figure=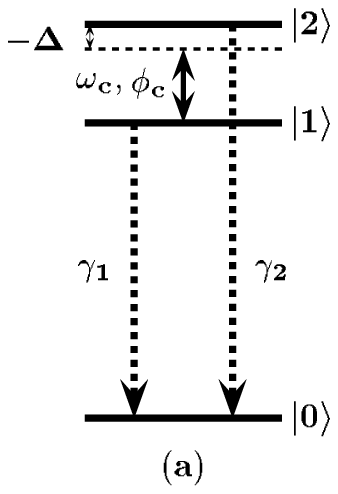,height=6cm}}}

\vspace*{2.cm}

\centerline{\hbox{\psfig{figure=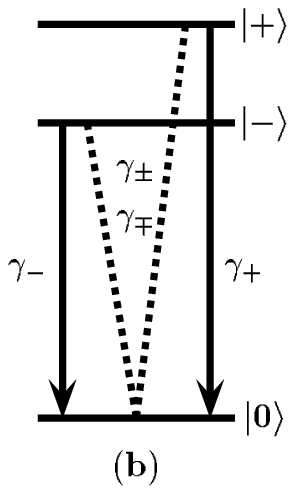,height=6cm}}}

\caption
{ The system under consideration: (a) bare states representation: The two excited states $|1 \rangle$, 
$|2 \rangle$ are coupled by a 
microwave field with frequency $\omega_{c}$ and phase $\phi_{c}$ and both decay spontaneously to a 
common ground state $|0 \rangle$.
(b) dressed states representation with dressed decay rates to the ground state $\gamma_+$ and $\gamma_-$ (solid lines)
and relaxation among the dressed states $\gamma_{\mp}$ from $|-\rangle$ to $|+\rangle$ and $\gamma_{\pm}$ 
in the opposite direction (dashed lines), which arises from a coupling of the 
transitions of both dressed states to the ground state. }
\end{figure}

\pagebreak
\begin{figure}
\centerline{\hbox{
\psfig{figure=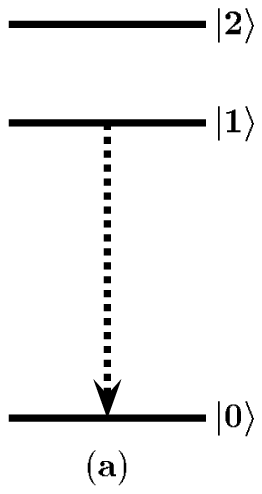,height=6.5cm} \hspace*{2.cm}
\psfig{figure=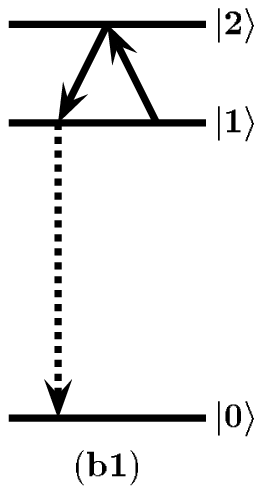,height=6.5cm}}}

\vspace*{2.cm}

\centerline{\hbox{
\psfig{figure=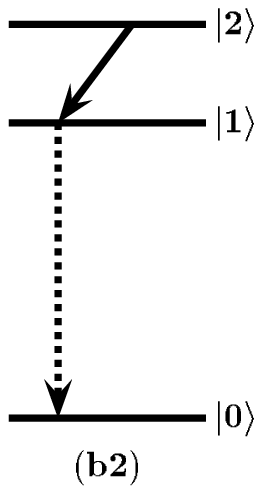,height=6.5cm} \hspace*{2.cm}
\psfig{figure=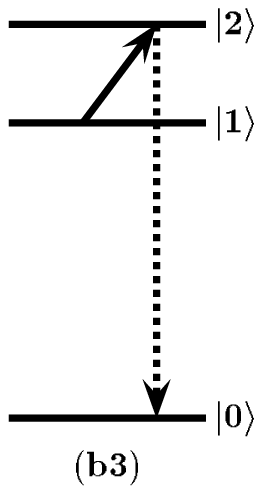,height=6.5cm}}}

\caption
{The leading
interfering paths for the decay process from e.g.\ the lower excited state; with solid lines we present the pathways for the driving field transitions and with dashed lines the spontaneous emission transitions.
(a) Simple spontaneous decay. (b1) Phase independent combined absorptions and
emissions of driving field photons via the higher excited state. (b2) Phase
dependent interfering paths for initial coherent superpositions via single
stimulated emission  from the higher excited state prior spontaneous emission.
(b3) Phase dependent single absorption of driving field photon followed by
spontaneous emission from the higher excited state. In this case the paths are
indistinguishable if the excited levels are closely spaced compared with the
decay rates and if both dipole matrix elements coupling to the ground state are parallel.}
\end{figure}
\pagebreak

\begin{figure}
\centerline{\hbox{\psfig{figure=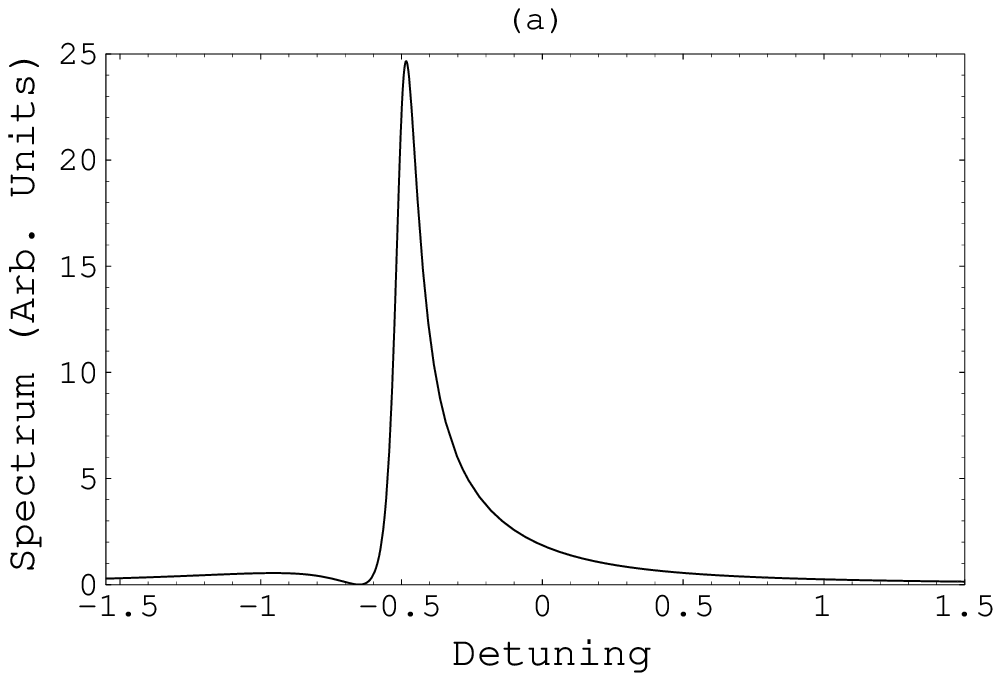,height=6.cm}}}
\centerline{\hbox{\psfig{figure=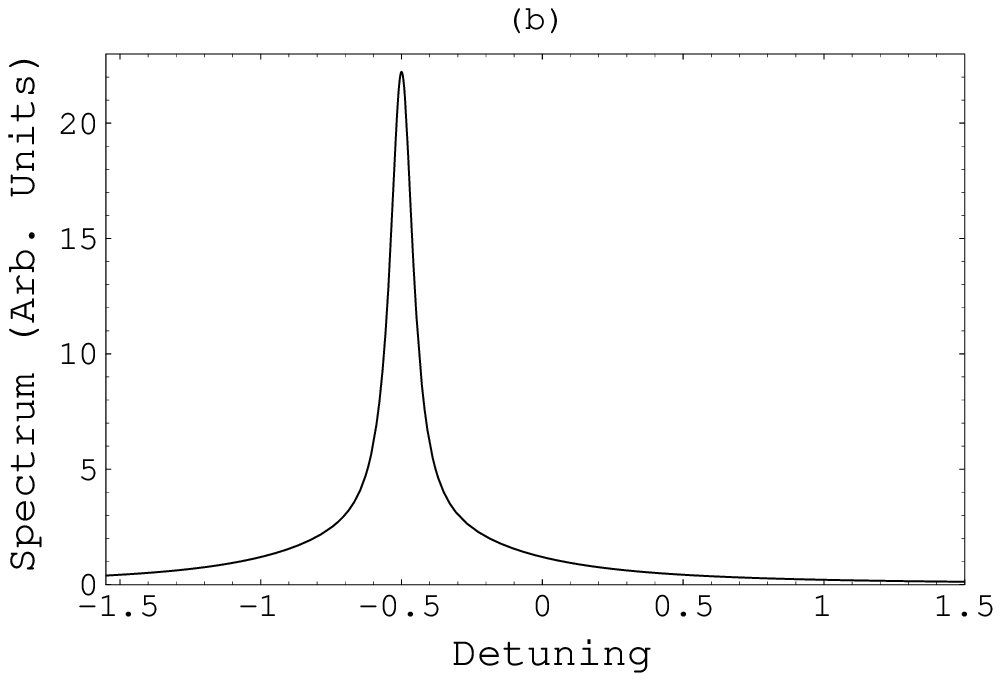,height=6.cm}}}
\centerline{\hbox{\psfig{figure=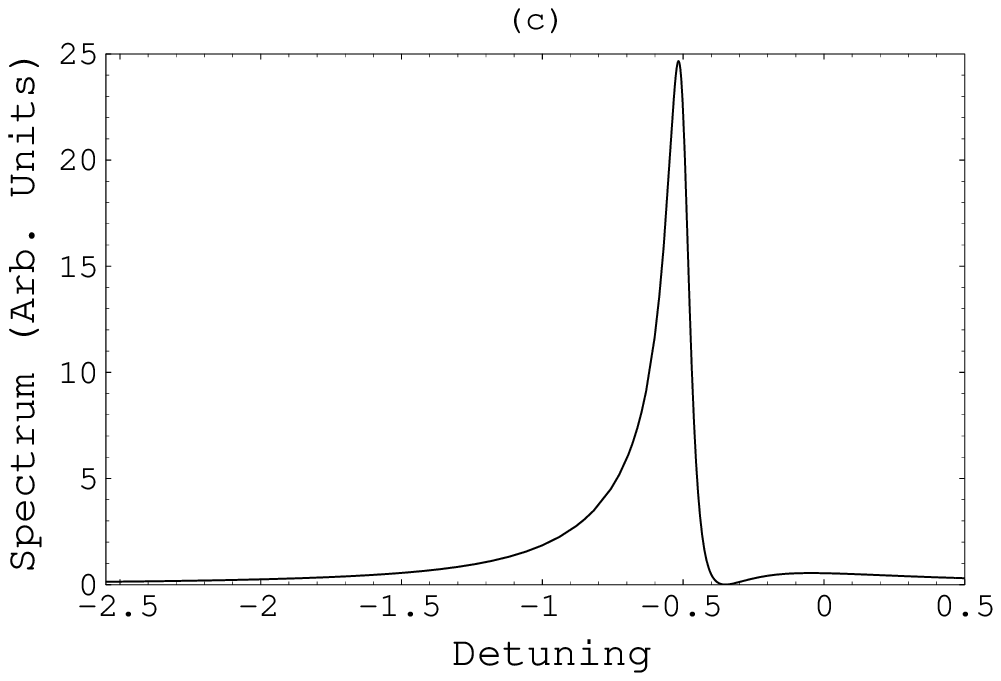,height=6.cm}}}
\caption
{The spontaneous emission spectrum $S(\delta_{\bf k})$
(in arbitrary units) for atomic parameters  $\gamma_{2} = 0,  \theta = \pi/4 $,
$\Omega = 0.15 \gamma_{1}$, $\Delta = 0$ and $\omega_{21} = \gamma_{1}$. In (a)
$\delta\phi = 0$, (b) $\delta\phi = 0.5\pi$ and (c) $\delta\phi = \pi$. 
The detuning $\delta_{\bf k}$ is measured in units of $\gamma_{1}$.}

\end{figure}

\pagebreak

\begin{figure}
\centerline{\hbox{\psfig{figure=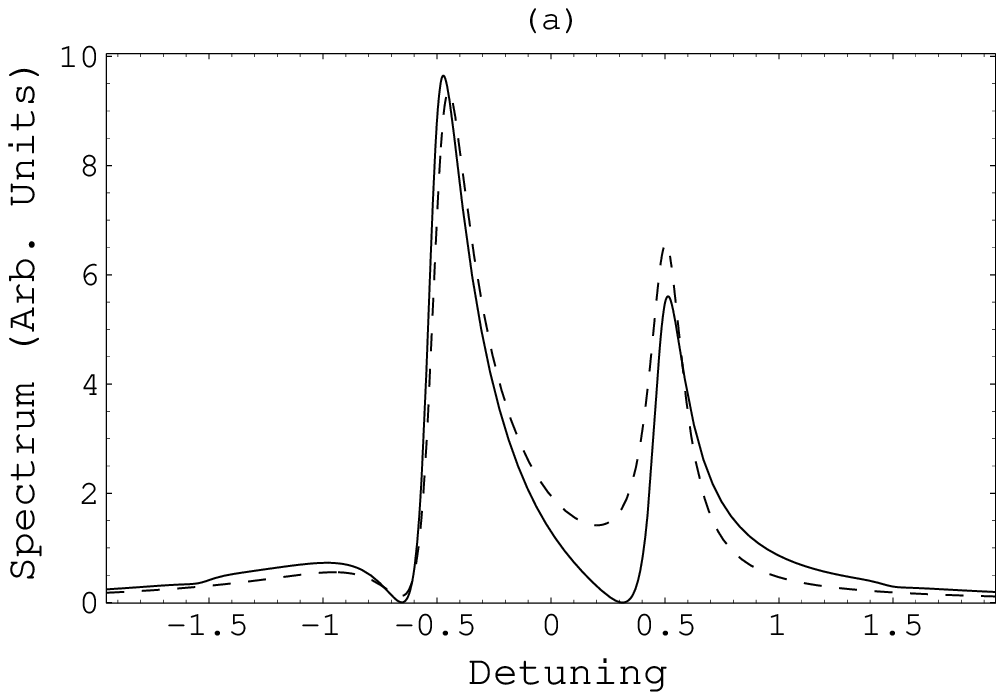,height=6.cm}}}
\centerline{\hbox{\psfig{figure=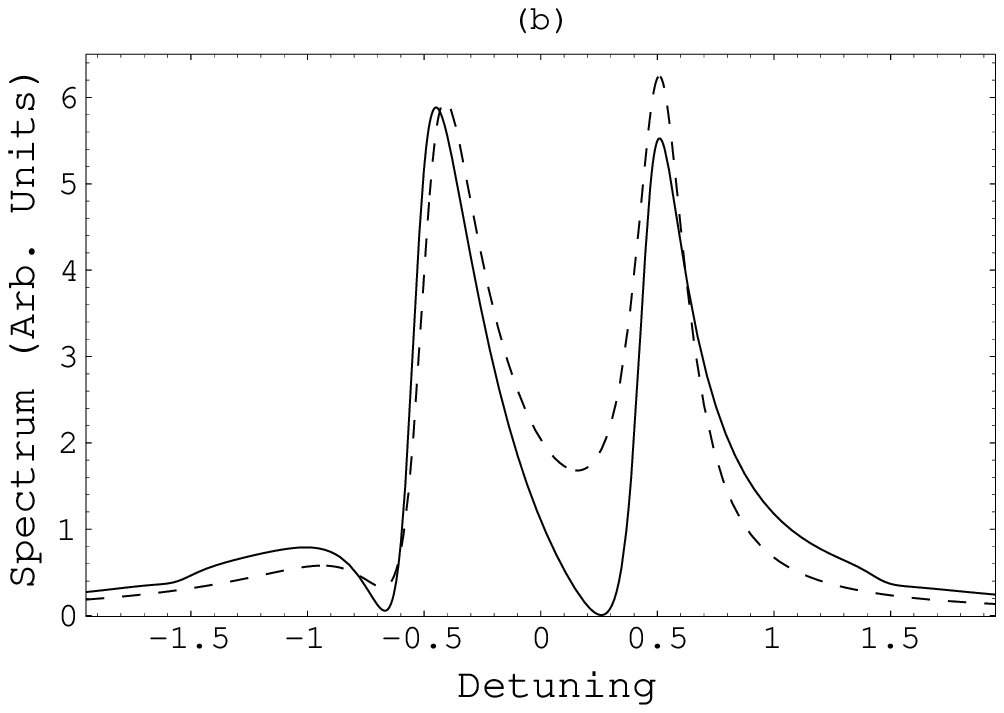,height=6.cm}}}
\centerline{\hbox{\psfig{figure=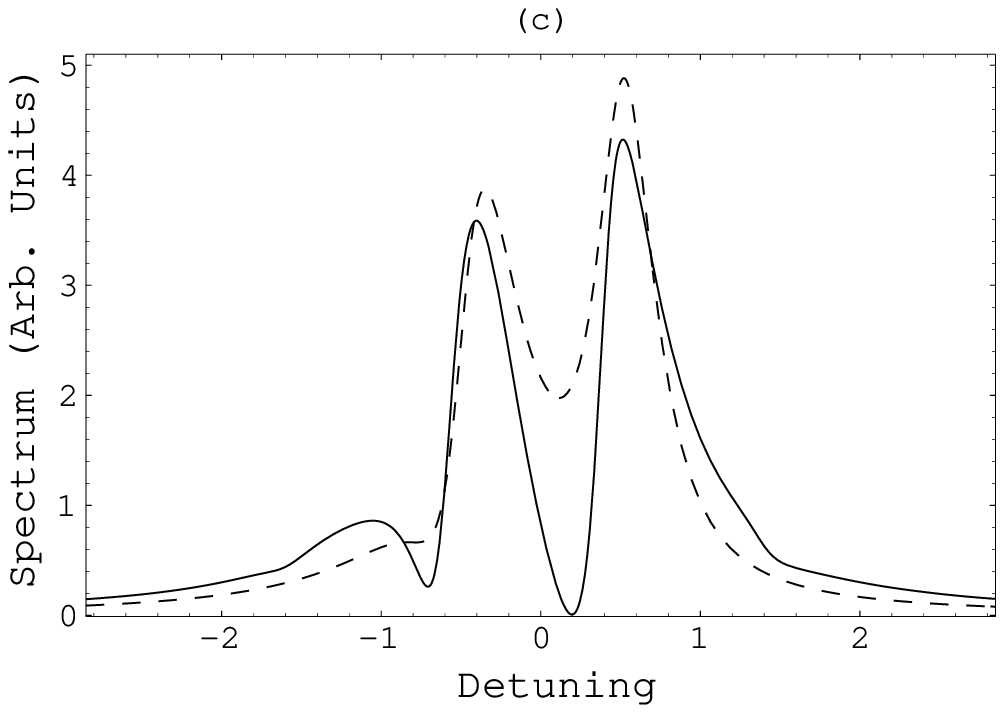,height=6.cm}}}

\caption
{The same as in figure 3(a) but for (a) $\gamma_{2} =
0.075 \gamma_{1} = 0.5 \Omega$,   (b) $\gamma_{2} = 0.15 \gamma_{1} = \Omega$
and (c) $\gamma_{2} = 0.3 \gamma_{1} = 2 \Omega$. For the dashed curve we present the results with $p = 0$ and for the solid curve the results with $p = 1$. The detuning $\delta_{\bf k}$ is measured in units of $\gamma_{1}$.}
\end{figure}

\begin{figure}
\centerline{\hbox{
\psfig{figure=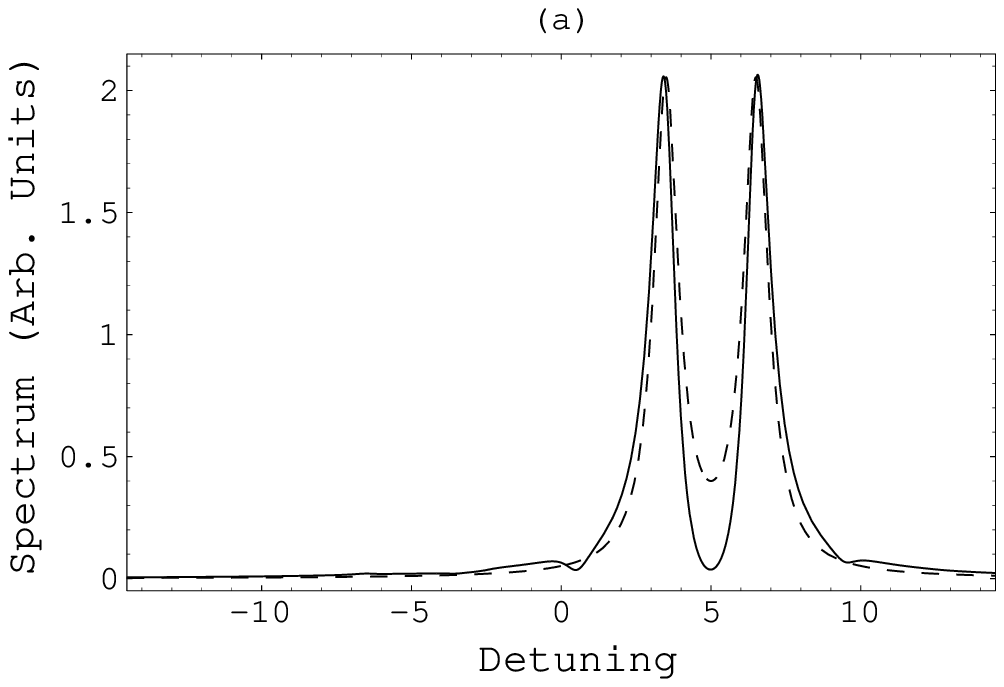,height=6.cm}
\psfig{figure=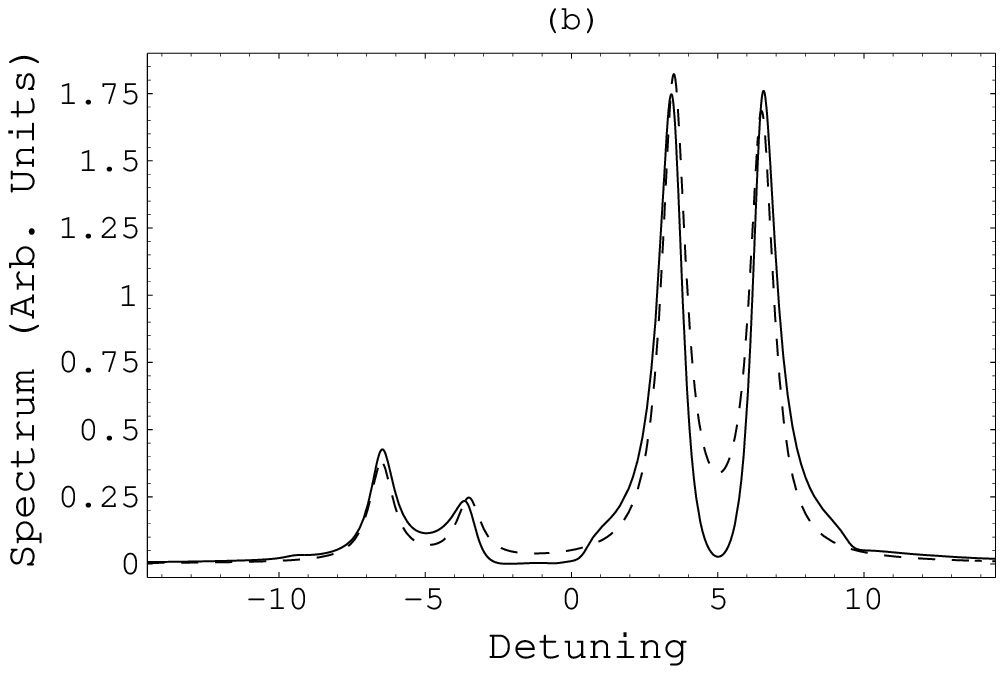,height=6.cm}}}
\centerline{\hbox{
\psfig{figure=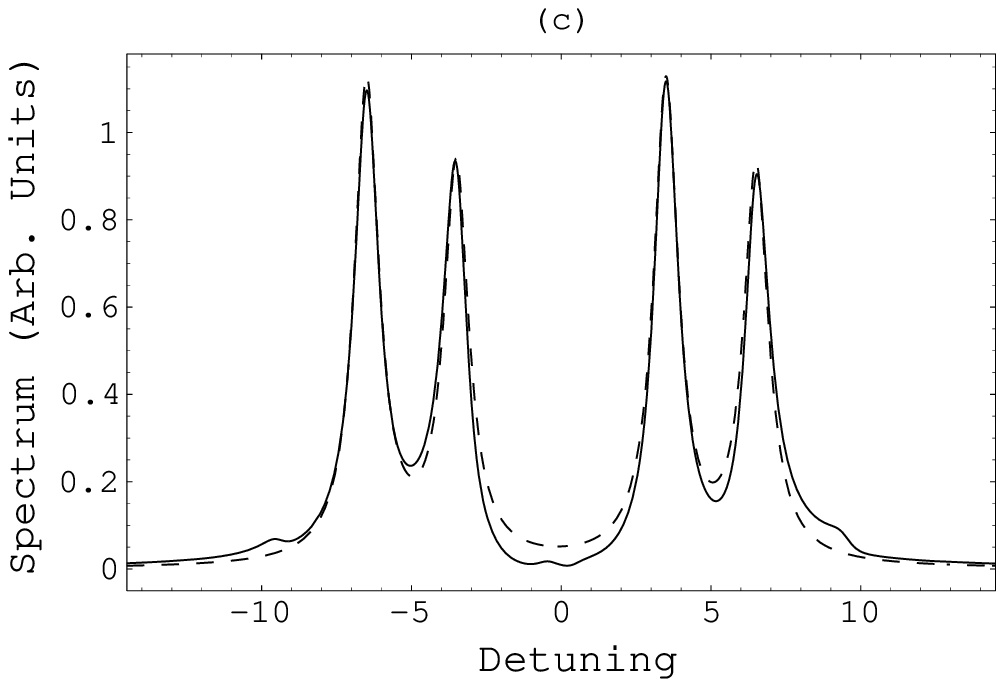,height=6.cm}
\psfig{figure=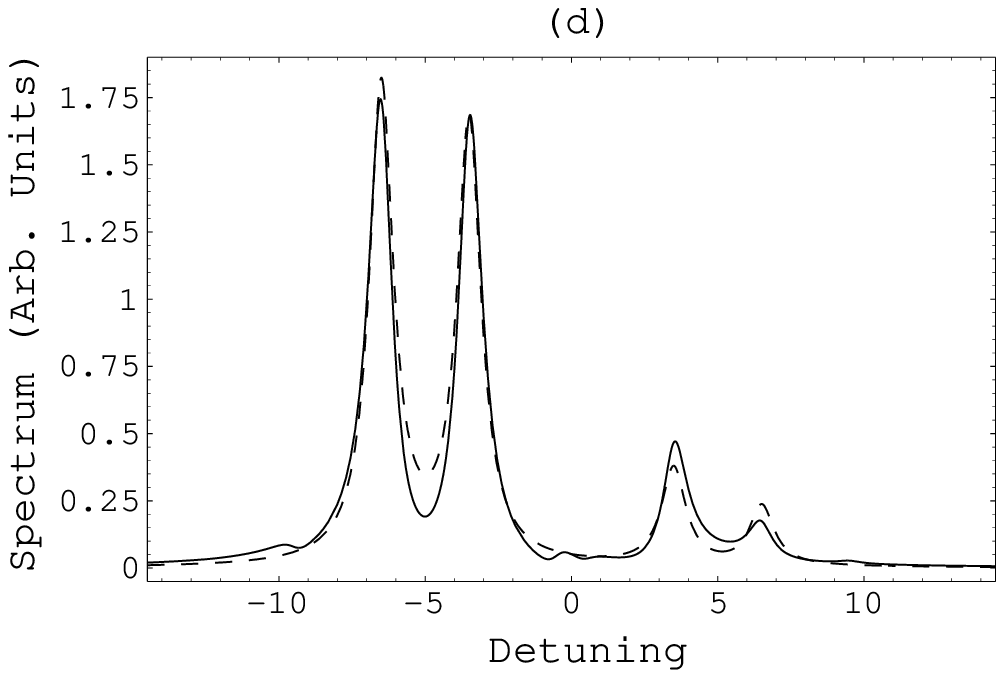,height=6.cm}}}
\centerline{\hbox{\psfig{figure=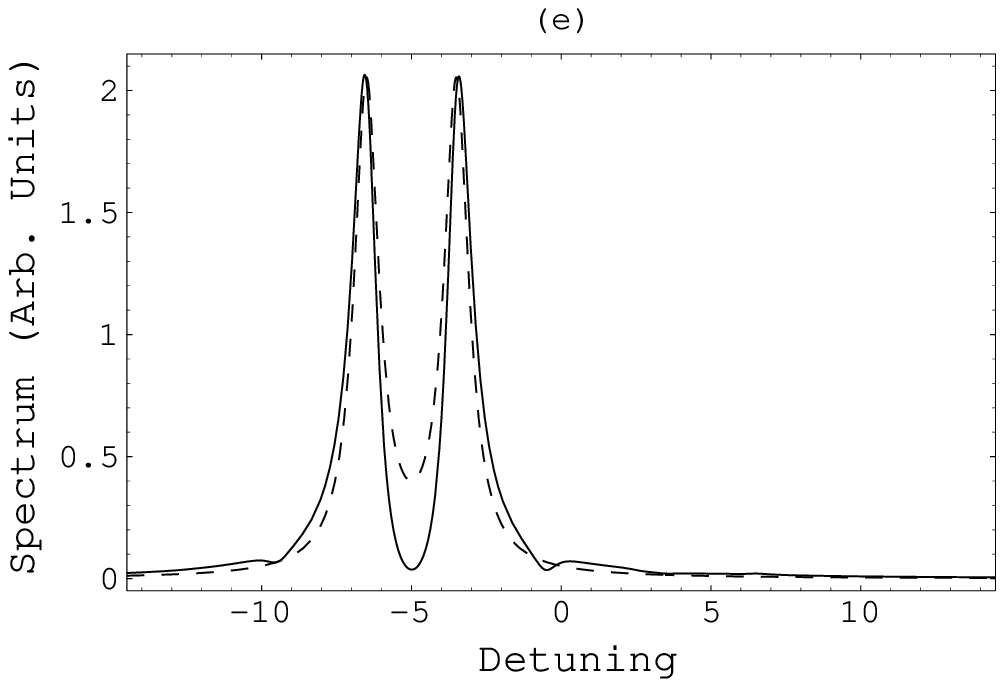,height=6.cm}}}
\caption
{The spontaneous emission spectra $S(\delta_{\bf k})$ (in arbitrary units) for atomic parameters $\gamma_{2} = \gamma_{1}$,
$\Omega = 5 \gamma_{1}$, $\Delta = 0$ and $\omega_{21} = 3 \gamma_{1}$. In (a) $\delta\phi = 0$, (b) $\delta\phi = 0.25\pi$, (c) $\delta\phi = 0.5 \pi$, (d) $\delta\phi = 0.75 \pi$ 
and (e) $\delta\phi = \pi$. For the dashed curve we present the results with $p = 0$ and for the solid curve the results with $p = 1$. The detuning $\delta_{\bf k}$ is measured in units of $\gamma_{1}$.
}
\end{figure}

\end{document}